	\newcommand{\be}{\begin{equation}}
        \newcommand{\ee}{\end{equation}}
        \newcommand{\ba}{\begin{eqnarray}}
        \newcommand{\ea}{\end{eqnarray}}
        \newcommand{\maps}{\colon}
        \newcommand{\BOX}{\hbox {$\sqcap$ \kern -1em $\sqcup$}}

        \newcommand{\qed}{\hskip 3em \hbox{\BOX} \vskip 2ex}
	\def\Stat{{\rm Stat}}
	\def\sgn{{\rm sgn}}
	\def\Maps{{\rm Maps}}

        \newcommand{\R}{{\bf R}}
        
        \newcommand{\Z}{{ \bf Z}}

	\newcommand{\id}{{\rm id}}
	\newtheorem{Theorem}{Theorem}
	\newtheorem{Corollary}{Corollary}
	\newtheorem{Lemma}{Lemma}
\def\a{ \alpha }
\def\b{ \beta }

\def\i{ \iota }

\def\S{\Sigma}

\def\({ \left( }
\def\){ \right) }
\def\[{ \left[ }
\def\]{ \right] }
\def\<{ \langle }
\def\>{ \rangle }

\def\^{ \wedge }
\def\X{ \times }
\def\V{ \vee }
\def\+{\oplus}

\def\.{ \cdot}
\def\@{ \otimes }

\def\:{\colon}

\def\Im{{\rm Im}}
\def\pinch{{\rm pinch}}

	\documentstyle[12pt]{article}
\begin{document}
        \begin{center}
      {\bf Topological Aspects of Spin and
	Statistics in Nonlinear Sigma Models\\}
        \vspace{0.5cm}
        {\em John C. Baez \\}
        \vspace{0.3cm}
        {\small Department of Mathematics\\
        Wellesley College \\
        Wellesley, Massachusetts 02181\\
(on leave from the University of California at Riverside)\\}
	\vspace{0.5cm}
	{\em Michael S. Ody \\}
        \vspace{0.3cm}
        {\small The Physics Laboratory\\
        University of Kent at Canterbury \\
        Kent, CT2 7NZ\\
	United Kingdom\\}
        \vspace{0.5cm}
        {\em William Richter \\}
        \vspace{0.3cm}
        {\small Department of Mathematics\\
        Massachusetts Institute of Technology \\
       Cambridge, Massachusetts 02139\\}
	\vspace{0.5cm}
	{\small May 14, 1992}

        \end{center}
\vbox{\vspace{1.2cm}}
\begin{abstract}
We study the purely topological restrictions on allowed spin and
statistics of topological solitons in nonlinear sigma models.
Taking as space the connected $d$-manifold
$X$, and considering nonlinear sigma models with the connected
manifold $M$ as target space, topological solitons are given by
elements of $\pi_d(M)$.  Any topological soliton $\alpha \in \pi_d(M)$
determines a quotient $\Stat_n(X,\alpha)$ of the group of framed
braids on $X$, such that choices of allowed statistics for
solitons of type $\alpha$ are given by unitary representations of
$\Stat_n(X,\alpha)$ when $n$ solitons are present.
In particular, when $M = S^2$, as in the $O(3)$ nonlinear
sigma model with Hopf term, and $\alpha \in
\pi_2(S^2)$ is a generator, we compute that $\Stat_n(\R^2,\alpha) =
\Z$, while $\Stat_n(S^2,\alpha) = \Z_{2n}$.  It follows that phase
$\exp(i\theta)$ for interchanging two solitons of type $\alpha$ on
$S^2$ must satisfy the constraint $\theta = k\pi/n$, $k \in \Z$, when
$n$ such solitons are present. \end{abstract}

\section{Introduction}

	The behavior of quantum systems under the interchange of
identical particles goes under the name of ``statistics.''  In
elementary quantum mechanics, the interchange of identical particles
is assumed to affect the wavefunction of a many-particle system only
by the introduction of a phase, and interchange is modelled in terms
of the symmetric group, $S_n$.  Thus for a system of $n$ identical
particles one assumes that there is a character
$\chi \maps S_n \to U(1)$ describing the statistics.  For $n \ge 2$,
$S_n$ has two characters: either
$\chi(\sigma) = 1 $ for all $\sigma \in S_n$, or $\chi(\sigma) =
\sgn(\sigma)$.
Thus interchanging two particles must give rise to a phase of $1$ or
$-1$, corresponding to bosonic and fermionic statistics, respectively.
Statistics in which particle interchange gives rise to
higher-dimensional unitary representations of $S_n$ have also been
studied; these go under the name of ``parastatistics'' or ``nonabelian
statistics'' \cite{GM,DR}.

More recently, in the physics of two-dimensional systems, it has been
noted that a more precise description of the interchange of identical
particles should keep track of the braid traced out by particles as
they are interchanged \cite{WilczekWu}.  For structureless point
particles on the plane this amounts to replacing the symmetric group
by the Artin braid group, $B_n$.  The group $B_n$ has a circle's worth
of characters; that is, interchanging two particles gives rise to an
arbitrary phase
$\exp(i\theta) \in U(1)$.  These statistics are known as ``anyonic'' or
``fractional.''  Anyonic statistics, as well as nonabelian anyonic
statistics, appear to play a crucial role in the fractional quantum
Hall effect \cite{MRPGXGWen,ShWi}.  Anyons have also been proposed
as a mechanism for superconductivity \cite{Wilczek2}.

There are further variations on the theme of statistics which we would
like to address here.
If one considers structureless point particles on the space $X$,
the proper analog of the Artin braid
group is the group $B_n(X)$ of the braids on
$X$ (defined below).    The implications for
statistics have been studied already by a number of authors
\cite{TW,IIS}.   For particles with internal structure,
however, one must also keep track of the rotation of the particles as
one interchanges them.   Thus it is natural to work with the
group $FB_n(X)$ of {\it framed} braids on $X$.
 This has the braid group of $X$ as a quotient group, that
is, there is a surjective homomorphism
$$
       \pi \maps FB_n(X) \to B_n(X) .
$$
A unitary representation of $FB_n(X)$ corresponds to a
choice not only of statistics, but also certain a certain
aspect of spin, namely, the phase induced by rotating a particle by
$2\pi$. In the case $X =
\R^2$, spin and statistics may be chosen independently, at least at
the level of many-particle quantum mechanics.  The mathematical reason
is that {\it in this case} there is a homomorphism
$$
    \iota  \maps B_n(X) \to FB_n(X)
$$
such that $\pi (\iota (g)) = g$ for all $g \in B_n(X)$.  In general,
however---for example, when $X = S^2$---this is {\it not} true, so
certain relations between spin and statistics arise.  These
are not to be confused with the spin-statistics relations arising in
quantum field theory.

For nonlinear sigma models, further constraints exist on the spin and
statistics of topological solitons.   This has been noted for the
2+1-dimensional $O(3)$ nonlinear sigma model with Hopf term by Wilczek
and Zee \cite{WZ}, and subsequently elaborated on by Wu and Zee
\cite{WuZ},
 Wen \cite{JWen}, and others.   Here we describe a general framework
to study fields from spacetime, $\R \times X$, to a target manifold
$M$.  (In the $O(3)$ nonlinear sigma model, $M = S^2$.)
Topological solitons correspond to elements of $\pi_d(M)$, where $d$
is the dimension of $X$.  Any topological soliton $\alpha$ defines a
certain quotient group $\Stat_n(X,\alpha)$ of $FB_n(X)$, and a
representation $\rho$ of $FB_n(X)$ corresponds to an allowed choice of
spin and statistics for solitons of type $\alpha$ if and only if
$\rho$ factors through $\Stat_n(X,\alpha)$, that is, if there exists a
representation $\widetilde \rho$ of $\Stat_n(X,\alpha)$ such that
$$           \rho = \widetilde\rho j $$
where $j \maps FB_n(X)\to \Stat_n(X,\alpha)$ is the quotient map.

An analysis along these lines leads to some interesting
results.  For example, using a braid group analysis Thouless
and Wu \cite{TW} concluded that the phase $\exp(i \theta)$ from
interchanging two particles on $S^2$ must satisfy
$$
      \theta = {k\pi\over(n-1)}, \qquad k \in \Z
$$
when $n$ particles are present.  This is correct for structureless
point particles.  However, a study of the $O(3)$
nonlinear sigma model gives a different result for
solitons with unit topological charge on $S^2$, namely:
$$
        \theta = {k\pi\over n} ,\qquad k\in \Z.
$$

As alluded to above, in quantum field theory the axioms of
locality, Poincar\'e-invariance, energy positivity, and so on give
rise to additional relationships between spin and statistics.
These are of a different character than our results, which are derived
in the context of many-particle quantum mechanics by purely topological
methods.  For spin-statistics theorems in quantum field theory,
we refer the reader to the original arguments of Fierz and Pauli
\cite{FierzPauli}, the theorems proved using the Garding-Wightman
\cite{SW} and C*-algebraic axioms \cite{DR,DHR}, and the more recent
extensions to field theory in 2 or 3 dimensions, in which anyonic
statistics arise \cite{FMFGFGMRS}.

\section{Spin, Statistics, and Framed Braids}

The relations between spin, statistics, and braid groups treated
 here arise naturally from considering
quantization of systems for which the classical configuration space
$C$ is not simply connected \cite{LMSo}.   While the most obvious
choice of the quantum Hilbert space is simply the space $L^2(C)$
of square-integrable functions on $C$, one may equally well
use $L^2(C,E)$, the space of $L^2$ sections of a flat line bundle $E$
over $C$.  Isomorphism classes of flat
line bundles over $C$ are in one-to-one correspondence with
characters of the fundamental group $\pi_1(C)$.  For any
group $G$, let $G^\ast$ denote the group of characters, or
one-dimensional unitary representations, of $G$.  Then each element of
$\pi_1(C)^\ast$ gives a different quantization.  (Of course, if $C$ is
infinite-dimensional there are severe analytical difficulties in
defining the appropriate $L^2$ spaces, which we do not address here.)

For a system of $n$ indistinguishable
structureless point particles on a connected manifold $X$, where we
assume that no two particles can be at the same place at the same
time, the configuration space is
$(X^n -  \Delta)/S_n,$ where $\Delta \subseteq X$ is given by
$$
     \Delta = \lbrace (x_1, \dots, x_n) \colon\;
\exists j\ne k \;\; x_j = x_k  \rbrace ,
$$
and the symmetric group $S_n$ acts on $X^n - \Delta$ by permutation of
the points $(x_1, \dots, x_n)$.   Thus quantizing this system involves
choosing an character of the {\it braid group of} $X$,
$$
          B_n(X) =   \pi_1((X^n -  \Delta)/S_n) .
$$
In other words elements of $B_n(X)^\ast$ correspond to choices of
(abelian) statistics.  Similarly, $k$-dimensional unitary
representations of $B_n(X)$ correspond to nonabelian statistics, as
they define  flat $U(k)$-bundles over the configuration space $(X^n -
\Delta)/S_n$.

If $\dim X > 2$, the braid group $B_n(X)$ equals $S_n$
when $X$ is simply connected \cite{Birman1}, and in general is the
wreath product of $S_n$ and $\pi_1(X)$ \cite{IIS}.     The
two-dimensional case is more interesting; for example, $B_n(\R^2) =
B_n$ is the original braid group due to Artin \cite{ArtinBirman2},
with generators $s_j$,  $1 \le j < n$, and relations
\ba       s_j s_k &=& s_k s_j   \qquad\qquad  |j - k| \ge 2 ,
\nonumber\cr
       s_j s_{j+1} s_j  &=& s_{j+1} s_j s_{j+1} .  \nonumber\ea
Here $s_j$ corresponds to the interchange of the $j$th and $(j+1)$st
particle in a counterclockwise manner.
Later Fadell and Van Buskirk computed $B_n(S^2)$
\cite{FvB}, Van Buskirk computed $B_n(RP^2)$ \cite{vB}, Birman
computed $B_n(T^2)$ \cite{Birman1}, and finally Scott
computed $B_n(X)$ for $X$ any compact 2-manifold \cite{Scott}.
For example, $B_n(S^2)$ is the quotient of $B_n$ by
the additional relation
\ba    s_1s_2 \dots s_{n-1}s_{n-1}s_{n-2} \dots s_1 = 1.
\label{sphere}\ea
This is not surprising, since any element of $B_n(S^2)$ arises from
a braid in $\R^2$, which has $S^2$ as its one-point compactification,
while the braid on the left side of equation (\ref{sphere})
corresponds to moving the first particle in a counterclockwise fashion
around the rest, and one may contract the loop traced out by the first
particle around the south pole of $S^2$ (the point at infinity).

It is easy to see that $B_n^\ast = U(1)$,
since all characters of $B_n$ are of the form
$$
       \chi(s_j) = e^{i\theta} .
$$
Equation \ref{sphere} implies that
$e^{2i(n-1)\theta} = 1$ for any character of $B_n$ that factors
through $B_n(S^2)$.  Thus $B_n(S^2)^\ast = \Z_{2(n-1)}$ for $n \ge 2$.
This was noted by Thouless and Wu \cite{TW}.

The situation changes when we consider particles with spin,
for example, topological solitons in a nonlinear
sigma model.  As we shall see, in this context one must treat spin and
statistics together using framed braids.  A framed braid may be
thought of as a ``ribbon'' \cite{RT}, but here we prefer to think of
it as a ``thickened'' braid.  Physically, a thickened braid represents
the world-tubes of a number of solitons.  Let $X$ be an oriented
connected manifold of dimension $d$, and let $D^d$ denote the closed
unit ball in $\R^d$.  Let
$e_i \maps D^d\to X$, $1 \le i \le n$, be disjoint oriented balls
embedded in $X$.  Let a {\it framed braid} on $X$ be an oriented
embedding $F$ of the disjoint union of $n$ solid cylinders
$[0,1] \times D^d$ in $[0,1]\times X$, such that
$F_i(0,\cdot) = F_i(1,\cdot) = e_i$, where $F_i$ denotes the embedding
of the $i$th cylinder, and such that
$$
F_i(t,x) = (t,F_{i,t}(x))
$$
for some function $F_{i,t}\maps D^d \to X$.   Let $FB_n(X)$ denote the
set of homotopy classes of framed braids on $X$, where the homotopy is
required to preserve the above conditions on $F$.     One can check
that $FB_n(X)$ is independent of the embeddings $e_i$, and there is a
canonical quotient map
$$
  \pi \maps FB_n(X) \to B_n(X) .
$$

The framed braid group keeps track not only of the interchange of
particles, but also their rotation in the process.  For example,
$FB_n = FB_n(\R^2)$ has generators $s_j$,
$1 \le j < n$, and $t_j$, $1 \le j \le n$, and relations
\ba
s_j s_k &=& s_k s_j   \qquad\qquad  |j - k| \ge 2,
\nonumber\cr
s_j s_{j+1} s_j &=& s_{j+1} s_j s_{j+1} ,
\nonumber\cr
s_j t_k &=& t_k s_j  \qquad\qquad    k \ne j, j+1 ,
\nonumber\cr
t_{j+1}s_j &=& s_j t_j ,
\nonumber\cr
t_j s_j &=& s_j t_{j+1} .
\nonumber\ea
The element $s_j$ corresponds to the interchange of the $j$th and
$(j+1)$st particles as before, while $t_j$ corresponds to a $2\pi$
rotation of the $j$th particle.   In this case there is a natural
inclusion $\iota \maps B_n \to  FB_n$ such that $\pi \circ \iota$ is
the identity on $B_n$.  In other words, the exact sequence
$$FB_n(X) \to  B_n(X) \to 1$$
 splits in this case; this occurs whenever the tangent bundle of $X$
is trivializable.
Thus every
character of $FB_n$ restricts to a character of $B_n$,
while conversely every character of $B_n$ extends to a
character of $FB_n$.  This allows us to
describe characters $\chi$ of $FB_n$ in terms of two independent
angles $\phi$ and $\theta$, related to spin and statistics,
respectively:
$$
       \chi(t_j) = e^{i \phi} ,\;\; \chi(s_j) = e^{i\theta} .
$$
Thus $FB_n^\ast = U(1) \times U(1)$ for $n \ge 2$.  (Note that the
character of $FB_n$ only depends on $\phi$ and $\theta$ modulo $2\pi$,
so it detects only the spin mod $\Z$ of the particle in question.)

On $S^2$, however, spin and
statistics are inextricably entangled, because $FB_n(S^2)$ is the
quotient of $FB_n$ by the relation
\ba    s_1s_2 \dots s_{n-1}s_{n-1}s_{n-2} \dots s_1 = t_1^2 .
\label{sphere2}\ea
In other words, the {\it framed} braid on $S^2$
 in which the first particle moves around the
rest but does not rotate about its own axis in the process is homotopic
to one in which the first particle experiences a rotation by $4\pi$.
This is easily visualized using a slight variant of the ``belt trick''
proof that $\pi_1(SO(3)) = \Z_2$, as in \cite{Kauffman}.
Thus there is no inclusion $\iota \maps B_n(S^2) \to FB_n(S^2)$ with
$\pi \circ \iota$ equal to the identity on $B_n(S^2)$, for $n \ge 2$.
Moreover, the character $\chi$ of $FB_n$ described above factors
through a character of $FB_n(S^2)$ if and only if $e^{2i(n-1)\theta} =
e^{2i\phi}$. It follows that $FB_n(S^2)^\ast \cong U(1) \times
\Z_{2(n-1)}$.

\section{Nonlinear Sigma Models}

In a nonlinear sigma model, fields are described as maps from
spacetime, $\R\times X$, to a target space $M$.   The
classical configuration space is thus the space of maps from $X$ to
$M$, denoted $\Maps(X,M)$.    The configuration space is a disjoint
union of connected components, one for each homotopy class in $[X,M]$.
One may construct a flat line bundle on $\Maps(X,M)$ from a flat line
bundle on each component.  A flat line bundle on the component of
$\Maps(X,M)$ containing a given map $f_0 \maps X \to M$ is uniquely
determined by a character on $\pi_1(\Maps(X,M), f_0)$.   Thus, as
described in the previous section, quantizing a given component of
$\Maps(X,M)$ depends on a choice of such a character.

We will study components of $[X,M]$ corresponding to collections of
topological solitons.   By a topological soliton, we mean
a map $g \maps X \to M$ that is constant outside a small ball in $X$.
With an appropriate Hamiltonian, solitons behave roughly like ``point
particles'' with internal degrees of freedom.
To work with topological solitons mathematically the Thom-Pontryagin
construction \cite{Milnor} turns out to be quite important.
Let $X$ be a compact oriented manifold of dimension $d$,
and let $M$ be a connected manifold with a chosen basepoint
$\ast$.  (We will discuss the case where $X$ is noncompact below.)
 Each element of $\pi_d(M)$ defines an element
$T(\alpha) \in [X,M]$ as follows.
Let $e \maps D^d \to X$ be any oriented embedded ball in $X$.
 Representing $\alpha$ by a map $g \maps D^d \to M$ with
$g|_{\partial D^d} = \ast$, we define $f_0 \maps X \to M$ by setting
$f_0 = g e^{-1}$ in the ball $e(D^d)$ and $f_0 = \ast$ outside
the ball.  The homotopy class of $f_0$ is obviously independent of the
choice of the representative $g \in \alpha$ and the choice of $e$,
so we may define the map
$$
      T \maps \pi_d(M) \to [X,M]
$$
by letting $T(\alpha) = [f_0]$.    Note that for any $n \ge 0$,
$T(n\alpha)$ can be represented by a map from $X$ to $M$ that is
constant outside of $n$ small balls, and looks like the map $f$ in
each ball.    Thus the path component of $\Maps(X,M)$ corresponding to
the element $TP(n\alpha)$ is the configuration space for the system
consisting of $n$ solitons of type $\alpha$.  Consequently, quantizing
this system requires a choice of a character of
$\pi_1(\Maps(X,M),f_0)$ where $f_0 \in \Maps(X,M)$ is any map with
homotopy class $T(n\alpha)$.

We now show how any choice of topological soliton $\alpha \in \pi_d(M)$
determines a quotient
group of the framed braid group $FB_n(X)$.   Representations of this
quotient group will correspond to allowed spin and statistics for
solitons of type $\alpha$.
First, choose a map $g \maps D^d \to M$ representing $\alpha$.
Then, given any element $[F] \in FB_n(X)$, recall that for $1 \le i
\le n$, $F_i \maps [0,1] \times D^d \to [0,1]\times X$ is an
embedding; physically the image of $F_i$ will represent the
world-tube of the $i$th soliton.   We write
$$
          F_i(t,x) = (t,F_{i,t}(x))
$$
where $(t,x) \in [0,1] \times D^d$.
We define a map $f \maps [0,1] \times X \to M$ by letting $f(t,x) =
g(F_{i,t}^{-1}(x))$ if $x \in X$ is in the image of $F_{i,t}$, and
$f(t,x) = \ast$ otherwise.  Note that $f(t,\cdot) = f_t$ is a loop of
maps from $X$ to $M$.    Thus $f$ defines an element $[f] \in
\pi_1(\Maps(X,M),f_0)$. Note also that $f_0 \in \Maps(X,M)$
has homotopy class $[f_0] = T(n\alpha)$.   Thus the group
$\pi_1(\Maps(X,M),f_0)$ only depends, up to isomorphism, on
$n$ and $\alpha$.    Moreover, it is
easy to check that $[f]$ is independent of the choice of $g$ with
$[g] = \alpha$ and the choice of framed braid $F \in [F]$.
It follows that there is a map
$$
          \psi \maps FB_n(X) \to \pi_1(\Maps(X,M),f_0) .
$$
We call the image of $\psi$ the {\it spin-statistics group} of $n$
topological solitons of type $\alpha$, and write this group as
$\Stat_n(X,\alpha)$.   It follows that the only allowed choices of
(abelian) statistics $\chi \in FB_n(X)^\ast$ for solitons of type
$\alpha$ are those which factor through
$\Stat_n(X,\alpha)$:
$$
\chi = \widetilde\chi \psi .
$$
Since $\psi$ is onto, abelian statistics are in fact in one-to-one
correspondence with elements of $\Stat_n(X,\alpha)^\ast$.  A similar
remark holds for nonabelian statistics, as mentioned in the
Introduction.

In the above we have assumed that space, $X$, is compact.  In the most
important noncompact case, $X = \R^n$, there would be no nontrivial
topological solitons according to the definition above, as all maps
$f_0 \maps \R^n \to M$ are homotopic.
In discussing solitons on $\R^n$ it is typical to work instead with its
compactification, $S^n$.  The reason usually given (which is admittedly
somewhat heuristic), is that
for non-topological terms in the action
for the field $f \maps \R \times \R^n \to M$ to be finite,
$f$ must be static
at spatial infinity; that is, for some point $\ast \in M$,
\begin{equation}
 \lim_{|\vec x| \to \infty} f(t,\vec x)  = \ast \in M
\label{static}
\end{equation}
for all times $t$.  Thus for each $t$, the map $f_t \maps \R^n \to M$
extends uniquely to a continuous map from $S^n$ to $M$ sending the
point at infinity to $\ast$.  In other words, the physically relevant
configuration space is the space $\Maps_\ast(S^n,M)$ of {\it basepoint
preserving} maps from $S^n$ to $M$, where we take the point at
infinity as the basepoint for $S^n$.  The rest of the analysis of spin
and statistics of solitons need be only slightly changed in order to
take this into account.

In general, when $X$ is noncompact, let $\overline X$ be the
one-point compactification of $X$, with the point at infinity as
basepoint.  Then there is a Thom-Pontryagin map
$$
T \maps \pi_d(M) \to [\overline X,M]_\ast ,
$$
where $[\overline X,M]_\ast$ denotes the basepoint preserving homotopy
classes of basepoint preserving maps from $\overline X$ to $M$.  Let
$\alpha \in \pi_d(M)$.
Then for any $f_0 \maps \overline X \to M$ with $[f_0] = T(n\alpha)$,
there is a map
$$
       \psi\maps FB_n(X) \to \pi_1(\Maps_\ast(X,M),f_0) ,
$$
defined just as in the compact case, and we define the image of $\psi$
to be $Stat_n(X,\alpha)$.   (In certain cases compactifications other
than the one-point compactification would be more appropriate, but the
necessary adjustments are easily made.)

We should note the mathematical resemblance between our techniques and
those arising in the use of ``configuration space models'' in homotopy
theory for computing the homology of iterated loop spaces $\Omega^n
\Sigma^n M$ \cite{Segal}.  In particular it is interesting to note the
use of creation and annihilation operators for topological solitons
\cite{CarMcD} in a purely mathematical context.

\section{Examples}

One nonlinear sigma model for which spin and statistics has been
deeply studied is the $O(3)$ nonlinear sigma model with Hopf term
\cite{ShWi}.   Here fields are given by maps from
$\R \times \R^2$ to $S^2$.  In the study of this model it has been
common to compactify space-time to $S^3$.
This technique eliminates from
consideration all fields with nonzero soliton number, a deficiency
which seems so far
only to have been addressed by J.\ Wen \cite{JWen}, although certain
aspects of the mathematics seem to be foreshadowed by the work of
Ringwood and Woodward \cite{RW}.   The framework
developed above, of course, treats spin and statistics for arbitrary
soliton number.  In this section we calculate $\Stat_n(\R^2,\alpha)$
and $\Stat_n(S^2,\alpha)$ for the soliton $\alpha$ corresponding to
the map of degree one from $S^2$ to itself.   In doing so we develop
techniques applicable to general simply-connected target manifolds
$M$.  In particular, we compute the groups
$$
\pi_1(\Maps(S^2,M),f_0)
$$
and
$$
\pi_1(\Maps_\ast(S^2,M),f_0)
$$
for any $f_0 \maps S^2 \to M$.

In what follows, we work in the category of spaces with basepoint,
so ``map'' will mean ``basepoint preserving map.''  Let $M$
be a connected and simply connected space with basepoint.
First, note that a loop in $\Maps_\ast(S^2,M)$ is the same as a
map $f \maps S^1 \times S^2/S^1 \times\ast$ to $M$.  Let
$$
\iota \maps S^2 \to S^1 \times S^2/S^1 \times \ast
$$
denote the natural inclusion.   This map induces a map
$$
\iota^\ast \maps [ S^1 \times S^2/S^1 \times \ast, M] \to \pi_2(M).
$$
In physical terms, $\iota^\ast[f] \in \pi_2(M)$ represents the
topological charge or ``soliton number'' of the field $f$.  In fact
the homotopy class of $f$ is completely determined by its soliton
number together with its ``instanton number,'' an element of
$\pi_3(M)$.

\begin{Theorem}\label{thm1}  For any simply connected space $M$,
$[ S^1 \times S^2/S^1 \times \ast, M]$ is isomorphic to $\pi_3(M)
\oplus \pi_2(M)$ in such a manner that the map
$$\iota^\ast\maps [ S^1 \times S^2/S^1 \times \ast, M] \to \pi_2(M) $$
corresponds to the projection $p_2 \maps \pi_3(M)
\oplus \pi_2(M) \to \pi_2(M)$.
\end{Theorem}

Proof -
Identify $S^3$ with the union of two solid tori,
$S^3 = D^2\X S^1 \cup_{S^1\X S^1} S^1\X D^2$.   Let us write a point
in $D^2\X S^1$ as $(t\vec x, \vec y)$ where $t \in [0,1]$ and $\vec
x, \vec y$ are unit vectors in $\R^2$, and similarly for $S^1\X D^2$.
Define the map $H \maps S^3 \to  S^1\X S^2 / S^1 \X *$ by
\ba
H(t\vec x, \vec y) &= *
\nonumber\cr
H(\vec x, t\vec y) &=   (\vec x,  \rho(t\vec y) ).
\nonumber\ea
Let $\V$ denote the wedge of spaces with basepoint.
The theorem is a consequence of the following lemma:

\begin{Lemma}
\label{Hopf-construction-gives-splitting}
The map $\i\V H \maps
S^2 \V S^3 \to S^1\X S^2 /  S^1 \X * $
is a homotopy equivalence.  Furthermore, $\pi_2 (\i\V H)$ and $\pi_1$
are homotopic as
maps from $S^2 \V S^3$ to $S^2$, and $\pinch (\i\V H)$ and $\pi_2$ are
homotopic as maps from $S^2 \V S^3$ to $S^3$, where
$\pi_1$ and $\pi_2$ are the natural quotient maps, and $\pinch \maps
S^1\X S^2 / S^1 \X * \to S^3$ is the pinch map.
\end{Lemma}

Proof - This follows easily from homology considerations.  \qed

\begin{Corollary}\label{cor1}  For any $f_0 \in \Maps_\ast(S^2,M)$,
$$
\pi_1(\Maps_\ast(S^2,M)),f_0) \cong \pi_3(M) .
$$
\end{Corollary}

Proof - This follows directly from the above theorem and the fact
that elements of $\pi_1(\Maps_\ast(S^2,M),f_0)$ are the
same as elements of $[S^1 \times S^2/S^1\times *,M]$ whose image under
$\iota^	\ast$ equals $[f_0] \in \pi_2(M)$.  \qed

In particular, to quantize the theory of maps $f \maps \R \times \R^2
\to S^2$ satisfying
$$
\lim_{|\vec x| \to \infty} f(t,\vec x) = \ast
$$
and having given soliton number
requires a choice of a representation of $\pi_3(S^2) = \Z$.  This
extends previous work that only treated the case of vanishing soliton
number \cite{WZ}.

The group
$\pi_1(\Maps(S^2,M), f_0)$ is a quotient of $\pi_1(\Maps_\ast(S^2,M),
f_0)$, and computing it requires an analysis of soliton-instanton
interactions.  The key topological aspects of these interactions are
are encoded in the Whitehead product
$$
         [\., \.] \maps \pi_2(M) \times \pi_2(M) \to \pi_3(M) .
$$
Let us briefly recall the definition of this product, referring the
reader to standard texts on algebraic topology for more details
\cite{Sp,Wh}.

The universal
Whitehead product is the map
$$
W\maps S^3 \to S^2 \V S^2
$$
given by
\ba
W(t\vec x, \vec y) &= \rho(t\vec x)\V *
\nonumber\cr
W(\vec x, t\vec y) &= *\V \rho(t\vec y).
\nonumber\ea
Given $\alpha = [f]$ and $\beta = [g]$ in $\pi_2(M)$, the Whitehead
product $[\alpha,\beta] \in \pi_3(M)$ is the class of the map $(f\V
g)\circ W$.  In particular, for any $\alpha \in \pi_2(M)$ the
Whitehead product defines a homomorphism
$$
[\alpha,\.] \maps \pi_2(M) \to \pi_3(M).
$$
In our application, $\alpha$ will be the class of $f_0 \maps S^2 \to
M$, so we write this homomorphism as $[f_0,\.]$.

The natural inclusion $\i\maps S^2 \to S^1\X S^2$
induces a map $\i^* \maps \[ S^1\X S^2, M \]\to \pi_2(M)$.
We have:

\begin{Theorem}
\label{Z2q}
For any simply connected space $M$ and any map $f_0\maps S^2 \to M$,
there is a bijection from the group $\pi_3(M)/ \Im[f_0,\.]$  to
the inverse image
$\i^*[f_0] \subseteq \[ S^2\X S^1, M \]$ of the homotopy class $[f_0]$.
\end{Theorem}

Proof - Let $\S A$ denote the suspension of a space $A$ and
$CA$ the cone over $A$.
In general for a cofibration sequence $A \hookrightarrow X \to X/A$
there is a left action of the group $[\S A, M]$ on the set $[X/A, M]$,
which has the property that there is a short exact sequence
\begin{equation}
\label{basic-action-exact-sequence}
   [X/A, M]/[\S A,M] \to [X, M]
		\to [A, M] .
\end{equation}
That is, there is an injective map from the orbit space
$ [X/A, M]/[\S A,M]$ to the set
$[X, M]$, whose image is exactly the elements which are mapped to the
trivial element in $[A,M]$ under $\i^*$.  The action is defined by the
coaction map
$\theta\maps  X\cup_\i CA \to X/A\V \S A$ together with the homotopy
equivalence $\gamma \maps X\cup_\i CA \to X/A$.  That is, given maps
$f\maps X/A \to M$ and $\a\maps \S A \to M$ there is a unique homotopy
class  $\a*f\maps X/A \to M$ such that the following two maps are
homotopic:
\begin{equation}
\label{definition-of-action-via-coaction}
(\a*f)\circ \gamma \sim (f \V\a)\circ \theta ,
\end{equation}
see  \cite{Sp}.

Now consider the cofibration sequence
$ S^1 \X * \subset S^1\X S^2 \to S^1\X S^2/ S^1\X *$.  Since
$\pi_1(M)  = 0$, diagram
(\ref{basic-action-exact-sequence}) implies that there is an
isomorphism
\begin{equation}
\label{exact-sequence-for-S2XS1toM}
 \[S^1\X S^2/ S^1\X * , M\]/\pi_2(M)
	\simeq \[S^1\X S^2, M\] .
\end{equation}
To compute the coaction map, let
$\tilde H\: S^3 \to S^1\X S^2 \cup C(S^1\times *)$ be the map
\ba
\tilde H(t\vec x, \vec y) &=& t\vec x \in D^2 = C(S^1 \times *)
\nonumber\cr
\tilde H(\vec x, t\vec y) &=& (\vec x, \rho(t\vec y))
\nonumber\ea
Then
\begin{equation}
\label{lift-equivalence-to-mapping-cone}
{i\V H} = (\i\V \tilde H)\circ j
\end{equation}
where
$$
      j \maps S^1\X S^2 \cup C(S^1\X *) \to S^1\X S^2 / S^1\X *
$$
is the natural homeomorphism.

The composite
\begin{equation}
\label{first-composite-for-theta}
(\pi_2\V\id)\circ \theta \circ \tilde H \maps S^3 \to S^2 \V S^2
\end{equation}
is clearly the universal Whitehead product $W\: S^3\to S^2 \V S^2$.

On the other hand the composite
\begin{equation}
\label{second-composite-for-theta}
(\pinch\V*) \circ \theta \circ (\i\V \tilde H) \maps S^2 \V S^3 \to
S^3
\end{equation}
is homotopic to the projection $\pi_2$ on the second factor.
Therefore by Lemma \ref{Hopf-construction-gives-splitting},
composites (\ref{first-composite-for-theta})
and (\ref{second-composite-for-theta}) and the Hilton-Milnor theorem
\cite{Wh},
\begin{equation}
\label{computation-of-theta}
(\i\V H\V \i) \circ (\i_1 \V ([\i_1,\i_3] + \i_2)) \sim
k \circ \i\V \tilde H
\end{equation}
as maps from $S^2 \V S^3$ to $S^1 \times S^2 /S^1 \times *$, where
$$ k \maps S^1 \times S^2 \cup C(S^1 \times *)
 \to S^1 \times S^2 /S^1 \times * $$
is the natural homotopy equivalence.
Lemma~\ref{Hopf-construction-gives-splitting} and
diagram (\ref{computation-of-theta}) computes the action of of
$\pi_2(M)$ on $\[S^1\X S^2/ S^1\X * , M\] \cong \pi_2(M) \+ \pi_3(M)$
to be
$$
\a*(f,\b) = (f, \b + [f,\a]).
$$
By the exact sequence (\ref{exact-sequence-for-S2XS1toM}) we have
$$
\pi_2(M) \+ \pi_3(M) \bigm/\lbrace (f,\b) = (f, \b + [f,\a])\rbrace
	\cong \[S^1\X S^2, M\] ,
$$
which implies Theorem~\ref{Z2q}.
\qed

\begin{Corollary}\label{cor2}
For any $f_0 \in \Maps(S^2,M)$,
$$
\pi_1(\Maps(S^2,M),f_0) \cong \pi_3(M)/\Im[f_0,\cdot]  .
$$
\end{Corollary}

Proof - This follows from the theorem above and the fact that
$\pi_1(\Maps(S^2,M),f_0) \cong \i^*[f_0] \subseteq [S^1 \X S^2,M]$.
\qed

We now compute the spin-statistics groups of the $O(3)$ nonlinear
sigma model:

\begin{Corollary}\label{cor3}  Let $\alpha \in \pi_2(S^2)$
 be a generator and let $f_0 \maps S^2 \to S^2$ have $[f_0] =
n\alpha$.  Then
$$
\pi_1(\Maps_\ast(S^2,S^2),f_0) = \Z, \qquad \pi_1(\Maps(S^2,S^2),f_0)
= \Z_{2n},$$
and for all $n \ge 1$,
$$
\Stat_n(\R^2,\alpha) = \Z, \qquad \Stat_n(S^2,\alpha) = \Z_{2n}.
$$
\end{Corollary}

Proof - Note that
$\pi_1(\Maps_\ast(S^2,S^2),f_0)$ is independent of $f_0$ and equals
$\pi_3(S^2) = \Z$ by Corollary \ref{cor1}.  Thus our choice of $\alpha$
determines a homomorphism $\psi \maps FB_n(\R^2) \to \Z$ as described
in Section 3.   Recall that $t_1 \in FB(\R^2)$ corresponds to the
rotation of the first strand by $2 \pi$ about its axis.
 We claim that $\psi(t_1) = 1$.  It will follow that
$\Stat_n(\R^2,\alpha)$, the image of the map $\psi$, is $\Z$.

Associate to the
framed braid $t_1$ a loop of basepoint preserving maps from $S^2$ to
$S^2$, i.e.\ a map $f \maps S^1 \times S^2 /S^1 \times * \to S^2$.  By
Lemma \ref{Hopf-construction-gives-splitting}, $\psi(t_1) \in
\pi_3(S^2)$ is represented by the map $f \circ H \maps S^3 \to S^2$.
We may calculate the Hopf invariant of this map by taking the inverse
images of two regular values in $S^2$ and computing their linking
number in $S^3$.   This is easily seen to be $1$, so $\psi(t_1) = 1$.

Next note that by Corollary \ref{cor2},
$\pi_1(\Maps(S^2,S^2), f_0)$ is the quotient of $\pi_3(S^2)$ by the
subgroup generated by $[\alpha,f_0]$.  Moreover
$[\alpha, f_0] = n[\alpha, \alpha] = 2n$ by the bilinearity of the
Whitehead product together with the fact that
$[\alpha,\alpha] = 2$.  Thus
$\pi(\Maps(S^2,S^2),f_0) = \Z_{2n}$.
Recall that $FB_n(S^2)$ is a quotient of $FB_n(\R^2)$.
Let $t \in FB_n(S^2)$ be the image of $t_1 \in FB(\R^2)$.   Then by the
same argument as for $t_1$, $\psi(t) = 1 \in \Z_{2n}$, so
$\Stat_n(S^2,\alpha) = \Z_{2n}$. \qed


\begin{thebibliography}{99}

\bibitem{GM} O.\ W.\ Greenberg and A.\ M.\ L.\ Messiah,
 Phys.\ Rev.\ {\bf
136}, B248 (1964); O.\ W.\ Greenberg and A.\ M.\ L.\ Messiah, Phys.\
Rev.\ {\bf 138}, B1155 (1965).

\bibitem{DR}   S.\ Doplicher and J.\ E.\ Roberts, in
{\sl Proc.\ of VIIIth Intl. Congress on
Math.\ Phys.}, eds.\ K.\ Mebkhout and R.\ S\'en\'eor (World
Scientific, New Jersey, 1987).

\bibitem{WilczekWu} F.\ Wilczek, Phys.\ Rev.\ Lett.\ {\bf 48}, 1144
(1982); F.\ Wilczek, Phys.\ Rev.\ Lett.\ {\bf 49}, 957 (1982);
 Y.-S.\ Wu, Phys.\ Rev.\ Lett.\ {\bf 52}, 2103 (1984);
Y.-S.\ Wu, Phys.\ Rev.\ Lett.\ {\bf 53}, 111 (1984).

\bibitem{ShWi}  A.\ Shapere and F.\ Wilczek, eds.,
{\sl Geometric Phases in Physics} (World Scientific, New Jersey,
1987).

\bibitem{MRPGXGWen}
R.\ E.\ Prange and S.\ M.\ Girvin, eds., {\sl The Quantum
Hall Effect} (Springer-Verlag, New York, 1987);
 G.\ Moore and N.\ Read, Nucl.\ Phys.\ {\bf B360} 362, (1991);
 X.-G.\ Wen, Phys.\ Rev.\ Lett.\ {\bf 66}, 802 (1991).

\bibitem{Wilczek2} F.\ Wilczek, ed., {\sl Fractional Statistics and
Anyon Superconductivity} (World Scientific, New Jersey, 1990).

\bibitem{TW}  D.\ J.\ Thouless and Y.-S. Wu,
 Phys.\ Rev.\  {\bf B31}, 1191 (1985).

\bibitem{IIS}
T.\ D.\  Imbo, C.\ S.\  Imbo and E.\ C.\ G.\ Sudarshan, Phys.\
Lett.\ {\bf B234}, 103 (1990).

\bibitem{WZ}  F.\ Wilczek and A.\ Zee, Phys.\ Rev.\ Lett.\ {\bf 51},
2250 (1983).

\bibitem{WuZ} Y.-S.\ Wu and A.\ Zee,  Phys.\ Lett.\ {\bf B147}, 325
(1984); Y.-S.\ Wu and A.\ Zee,  Nucl.\ Phys.\  {\bf B272} 322 (1986).

\bibitem{JWen} J.\  Wen, Commun.\ Theor.\ Phys.\ {\bf 16}, 111 (1991).

\bibitem{FierzPauli}  M.\ Fierz, Helvetica Physica Acta {\bf 12}, 3
(1939); W.\ Pauli, Phys.\ Rev.\ {\bf 58}, 716 (1940).

\bibitem{SW} R.\ F.\ Streater and A.\ S.\ Wightman, {\sl PCT, Spin,
Statistics and All That} (Addison-Wesley, New York, 1989).

\bibitem{DHR}  S.\ Doplicher, R.\ Haag and J.\ E.\ Roberts,
 Comm.\ Math.\ Phys.\ {\bf 23}, 199
(1971); S.\ Doplicher, R.\ Haag and J.\ E.\ Roberts,
 Comm.\ Math.\ Phys.\ {\bf 35}, 49 (1974).

\bibitem{FMFGFGMRS} J.\ Fr\"ohlich and P.\ A.\ Marchetti,
Lett.\ Math.\ Phys.\ {\bf 16}, 347 (1988);
 K.-H.\ Rehren and B.\ Shroer, Nuc.\ Phys.\ {\bf B312},
715 (1989); J.\ Fr\"ohlich, F.\ Gabbiani, and P.-A.\ Marchetti, in {\sl
Physics Geometry and Topology,} ed.\ H.\ C.\ Lee (Plenum, New York,
1991), p.\ 15; J.\ Fr\"ohlich and
F.\ Gabbiani, preprint.

\bibitem{LMSo} M.\ G.\ G.\ Laidlaw and C.\ Morette-DeWitt, Phys.\ Rev.\
 {\bf D3}, 1275 (1971); R.\ D.\ Sorkin, Phys.\ Rev. {\bf D27}, 1787
(1983).

\bibitem{ArtinBirman2} E.\ Artin, Ann.\ Math.\ {\bf 48}, 101 (1947);
J.\ Birman, {\sl Braids, Links, and Mapping Class
Groups} (Princeton U.\ Press, New Jersey, 1975).

\bibitem{Birman1} J.\ Birman, Comm.\ Pure.\ Appl.\ Math.\ {\bf 22}, 41
 (1969).

\bibitem{FvB} E.\ Faddell and J.\ Van Buskirk, Duke Math.\ Jour.\ {\bf
29}, 243 (1962).

\bibitem{vB} J.\ van Buskirk, Trans.\ Amer.\ Math.\ Soc.\ {\bf 122}, 81
(1966).

\bibitem{Scott} G.\ P.\ Scott, Proc.\ Camb.\ Phil.\ Soc.\ {\bf 68}, 605
(1970).

\bibitem{RT} N.\ Reshetikhin and V.\ Turaev, {\sl Comm.\ Math.\ Phys.\
}{\bf 127}, 1 (1990).

\bibitem{Kauffman} L.\ Kauffman, {\sl Knots and Physics} (World
Scientific, New Jersey, 1991).

\bibitem{Milnor} J.\ W.\ Milnor, {\sl Topology from the Differentiable
Viewpoint} (University Press of Virginia, Virginia, 1965).

\bibitem{Segal} G.\ B.\ Segal, Invent.\ Math.\ {\bf 21}, 213 (1973).

\bibitem{CarMcD}  J.\ L.\ Caruso,
Trans.\ Amer.\ Math.\ Soc.\ {\bf 265}, 147 (1981); D.\ McDuff,
Configuration spaces of positive and negative particles,
Top.\ {\bf 14}, 91 (1975).

\bibitem{RW}   G.\ A.\ Ringwood and L.\ M\ Woodward,
Phys.\ Rev.\ Lett.\ {\bf 53}, 1980 (1984).

\bibitem{Sp} E.\ H.\ Spanier, {\sl Algebraic Topology} (McGraw-Hill,
New York, 1966).

\bibitem{Wh} G.\ W.\ Whitehead, {\sl Elements of Homotopy Theory}
(Springer-Verlag, New York, 1978).

































\end{thebibliography}
\end{document}